# Towards ferroelectrically-controlled magnetism: Magnetoelectric effect in Fe/BaTiO$_3$ multilayers

Chun-Gang Duan, Sitaram S. Jaswal, and Evgeny Y. Tsymbal

*Department of Physics and Astronomy, Center for Materials Research and Analysis,
University of Nebraska, Lincoln, Nebraska 68588*

An unexplored physical mechanism which produces a magnetoelectric effect in ferroelectric/ferromagnetic multilayers is studied based on first-principles calculations. Its origin is a change in bonding at the ferroelectric/ferromagnet interface that alters the interface magnetization when the electric polarization reverses. Using Fe/BaTiO$_3$ multilayers as a representative model we show a sizable difference in magnetic moments of Fe and Ti atoms at the two interfaces dissimilar by the orientation of the local electric dipole moment. The predicted magnetoelectric effect is comparable in magnitude with that observed in elastically-coupled composites and opens a new direction to control magnetic properties of thin-film layered structures by electric fields.

Multiferroic materials are of great scientific and technological interest due to their magnetoelectric properties, originating from the coupling between ferroelectric and ferromagnetic order parameters.[1,2,3,4] The interplay between ferroelectricity and magnetism allows a magnetic control of ferroelectric properties[5] and an electric control of magnetic properties,[6] and could yield new device concepts, such as ferroelectric and multiferroic tunnel junctions.[7,8] Recently, it became possible to fabricate composite multiferroics by artificially making ferroelectrics and ferromagnets in nanoscale heterostructures. The two-phase multiferroics may play important role in future magnetoelectric devices because none of the existing single-phase multiferroics combine large and robust electric and magnetic polarizations at room temperature.[4] In a thin-film geometry, such composites can be created in two extreme forms: a multilayer consisting of alternating layers of the ferroelectric and ferro(ferri)magnetic phases or a vertically aligned columnar nanostructure. When the magnetoelectric coupling is exclusively caused by elastic interactions, the effect in a multilayer structure is expected to be negligible due to clamping to the substrate. On the other hand, in vertically aligned nanostructures the magnetoelectric effect may be significant, as was recently demonstrated for ferrimagnetic nanopillars embedded in a ferroelectric matrix.[9,10]

The coupling between elastic components of the ferromagnetic and ferroelectric constituents through the strain is not, however, the only source of a magnetoelectric effect in composite multiferroics. There is another physical mechanism that may cause ferroelectricity to influence magnetism and that may be sizable not only in vertical nanostructures but also in multilayers. It involves the coupling between ferroelectricity and magnetism through interface bonding. Displacements of atoms at the interface caused by ferroelectric instability alter the overlap between atomic orbitals at the interface which affects the interface magnetization. This produces a magnetoelectric effect the essence of which is the sudden change in the interface magnetization induced by the polarization reversal in the ferroelectric layer under the influence of applied electric field.

In this paper we explore the significance of the magnetoelectric effect driven by interface bonding in ferromagnetic/ferroelectric multilayers. We consider a Fe/BaTiO$_3$(100) multilayer as a representative composite multiferroic to investigate this phenomenon. We show a sizable difference in magnetic moments of Fe and Ti atoms at the two interfaces dissimilar by the orientation of the local electric dipole moment. The predicted magnetoelectric effect is compa-rable in magnitude with that observed in elastically-coupled composites and opens a new direction to control magnetic properties of thin-film layered structures by electric fields.

The motivation for this choice of materials is as follows. Fe and $BaTiO_3$ are two "classic" ferroic materials which have well-known properties in the bulk. Barium titanate is a perovskite ferroelectric oxide that has the spontaneous polarization $P_S \approx 0.26\,C/m^2$ and a Curie temperature $T_C \approx 393\,K$ at which the cubic–tetragonal transition occurs. Iron is a metallic ferromagnet that has a magnetic moment of $2.22 m_B$ per atom and a Curie temperature $T_C \approx 1043\,K$. Importantly, Fe (110) and $BaTiO_3$ (100) have a very good match of the lattice constants (a mismatch is only about 1.4%) that allows layer-by-layer epitaxial growth of Fe/$BaTiO_3$ miltilayers with no misfit dislocations.

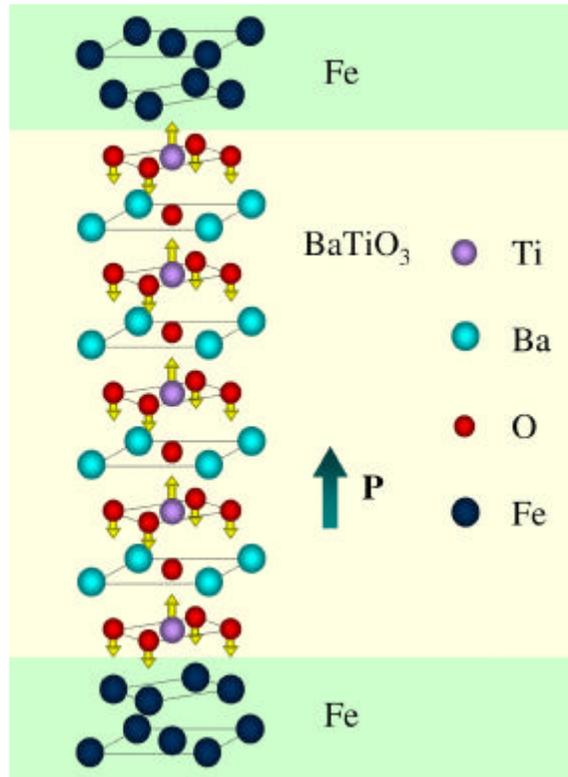

**Figure 1.** Atomic structure of Fe/$BaTiO_3$ multilayer for $m = 4$. Arrows indicate schematically displacements of Ti and O atoms in $BaTiO_3$ with the net polarization pointing up.

In order to elucidate the influence of ferroelectricity on magnetism, we have carried out density-functional calculations (VASP)[11] of the electronic and atomic structure of Fe/BaTiO$_3$(100) multilayers. The exchange-correlation potential is treated in the generalized gradient approximation (GGA). We use the energy cut-off of 500 eV for the plane wave expansion of the PAWs and a 10 x 10 x 1 Monkhorst Pack grid for *k*-point sampling. The convergences over both cut-off energy and *k*-point sampling have been tested. All the structural relaxations are performed until the Hellman-Feynman forces on atoms become less than 20 meV/Å. Local orbital-resolved densities of states and local magnetic moments are calculated by projecting the wave functions onto spherical harmonics centered on particular atoms as is implemented in VASP.

In the calculations we build up supercells by aligning the body centered cubic iron [110] axis and the [100] axis of the BaTiO$_3$. The in-plane lattice constant is fixed to be the experimental value of the bulk BaTiO$_3$ (3.991 Å), which is smaller than the theoretical lattice constant we calculated for the cubic phase of BaTiO$_3$ (4.033 Å). Hence, the in-plane ferroelectric instability is suppressed by the compressive strain.[12] This in-plane constraint is applied to relax the bulk structures of Fe and BaTiO$_3$. Under this constraint we find that the polarization of the bulk BaTiO$_3$ is 0.32 C/m$^2$, as calculated using Berry's phase method,[13] and the magnetic moment of the bulk Fe is 2.20 $\mu_B$ per atom. The obtained tetragonal structures are then used as building blocks for the Fe/BaTiO$_3$ supercells. Note that periodic boundary conditions of the supercell geometry impose the short-circuit condition between the Fe metal layers. To assure the accuracy of our calculations, we verified our results for a $m = 2$ multilayer against those computed using a more accurate full-potential linearized augmented plane-wave (FLAPW) method.[14]

We find that the most stable Fe/BaTiO$_3$ structure has a TiO$_2$ terminated interface with interfacial O atoms occupying atop sites on Fe which is similar to the result obtained previously for Co/SrTiO$_3$ multilayers.[15] Therefore supercells are constructed as (Fe$_2$)$_9$-TiO$_2$-(BaO-TiO$_2$)$_m$, where $m = 2, 4, 6, 8, 10$ and $16$. Fig. 1 shows the atomic structure of the $m = 4$ multilayer. First, we analyze properties of Fe/BaTiO$_3$ multilayers assuming that BaTiO$_3$ is in a paraelectric state. For this purpose we impose a mirror plane on the central TiO$_2$ layer and minimize the total energy of the whole system. We find that, although the net polarization of the BaTiO$_3$ film is zero, bonding at the interface induces interface dipole moments, which are oriented in the opposite directions at the two interfaces.[16]

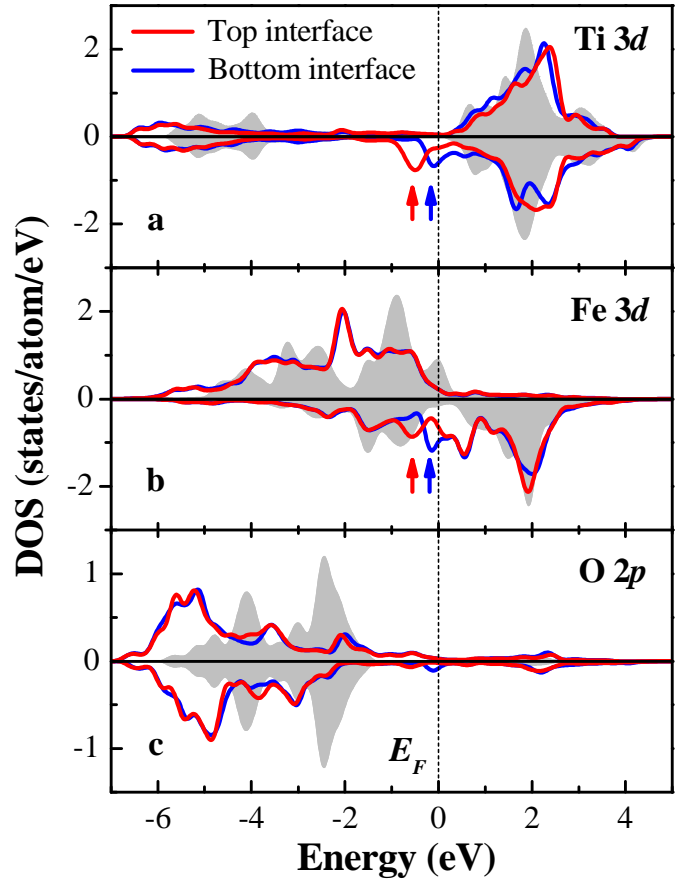

**Figure 2.** Orbital-resolved density of states (DOS) for interfacial atoms in a Fe/BaTiO$_3$ multilayer for $m = 4$. (a) Ti 3$d$; (b) Fe 3$d$; (c) O 2$p$. Majority-spin and minority-spin DOS are shown in upper and lower panels respectively. Red and blue curves correspond to the DOS of atoms at the top and bottom interfaces respectively. Shaded plots are the DOS of atoms in the central monolayer of Fe (b) or TiO$_2$ (a,c) which can be regarded as bulk. The vertical line indicates the Fermi energy ($E_F$).

Magnetic properties of the multilayer are due to ferromagnetism of Fe. In the paraelectric state, by symmetry, the magnetic moments of the interfacial atoms are exactly the same at the bottom and top interfaces. The calculated magnetic moment of the interface Fe atoms is notably enhanced up to about 2.64 $\mu_B$ compared to the bulk moment of 2.20 $\mu_B$. This enhancement is however not as significant as that for the Fe(100) surface, where the magnetic moment of the Fe surface atom is 2.98 $\mu_B$, indicating the involvement of bonding and charge transfer effects at the Fe/BaTiO$_3$ interface. The latter fact is reflected in the presence of induced magnetic moments on O and Ti atoms. We find that the magnetic moment of the interface O atom is about 0.05 $\mu_B$ and is parallel to the magnetic moment of Fe. On the other hand, the magnetic moment of Ti atom is about 0.30 $\mu_B$ and is antiparallel to the magnetic moment of Fe.

To reveal the influence of ferroelectricity on magnetism, we relax the constraint of reflection symmetry and minimize the total energy with respect to atomic coordinates of all atoms in the multilayer. For a $m = 2$ supercell, we obtain no ferroelectric instability, making all the results to be essentially the same as those in the presence of the symmetry constraint. Thickness $t \approx 1$ nm of the BaTiO$_3$ film corresponding to $m = 2$ appears to be the critical value for ferroelectricity. Increasing the thickness up to $t \approx 1.8$ nm ($m = 4$) is sufficient for ferroelectric instability to develop (Fig. 1).

Ferroelectric displacements break the symmetry between the top and bottom interfaces causing magnetic moments of Fe and Ti atoms at the two interfaces to deviate from their values in the paraelectric state. For the $m = 4$ multilayer the magnetic moment of Fe atoms at the bottom interface (the one at which the ferroelectric polarization is pointing away from the Fe layer as in Fig.1) is enhanced up to $\mu = 2.67$ $\mu_B$, while the magnetic moment of Fe atoms at the top interface (the one at which the ferroelectric polarization is pointing toward the Fe layer as in Fig.1) is reduced down to $\mu = 2.61$ $\mu_B$, which makes a difference in the Fe magnetic moments at the two interfaces $?\mu_{Fe} = 0.06$ $\mu_B$. A more sizable asymmetry is produced by ferroelectric displacements between the Ti magnetic moments at the two interfaces: $\mu = -0.40$ $\mu_B$ and $\mu = -0.18$ $\mu_B$ for the top and bottom interfaces respectively, so that the magnetic moment difference is $?\mu_{Ti} = 0.22$ $\mu_B$.

To understand the nature of the induced interface magnetic moments and the origin of the difference between the magnetic moments at the top and bottom interfaces we calculated orbital-resolved local densities of states (DOS). The results for the Ti 3$d$, Fe 3$d$, O 2$p$ orbitals are shown in Figs.2a-c indicating the presence of hybridizations between these states. Due to exchange splitting of the 3$d$ bands in Fe, these hybridizations produce exchange-split bonding and anti-bonding states which are the origin of the induced magnetic moments on the interface Ti and O atoms. The induced magnetic moment on the O atom is relatively small (0.05 $\mu_B$) because, as is seen from Fig.2c, the O 2$p$ orbitals lie well below the Fermi energy ($E_F$), and hence have a small overlap with the Fe 3$d$ states. The situation is however different for Ti atoms. The Ti 3$d$ band is centered at about 2 eV above the Fermi energy (the shaded plot in Fig.2a) and overlaps strongly with the minority-spin Fe 3*d* band which has a significant weight at these energies (the shaded plot in the lower panel of Fig.2b). The hybridization between the Fe and Ti 3*d* orbitals produce bonding states which are pushed down in energy and peaked just below $E_F$ (the peaks indicated by arrows in Figs.2a and 2b). Thus, the minority-spin Fe-Ti 3*d* bonding states cause charge redistribution between majority and minority spins resulting in a larger occupation of the minority-spin states of Ti. This implies an induced magnetic moment on Ti aligned *antiparallel* to the magnetic moment of Fe where majority-spin states have (by definition) greater occupation than minority-spin states.

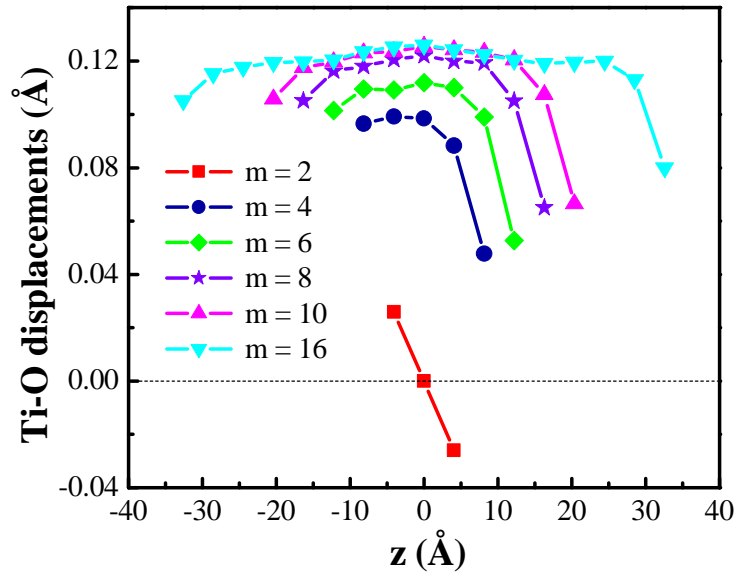

**Figure 3.** Relative Ti-O displacements in a ferroelectric BaTiO$_3$ film for different *m*. Positive values correspond to the local polarization pointing up as in Fig. 1. The midpoint of the ferroelectric layer lies at $z = 0$.

As we saw, the presence of ferroelectricity in BaTiO$_3$ causes the magnetic moments of interface Fe and Ti atoms to deviate from their values in the paraelectric state. This is due to the change in the strength of bonding between the Fe and Ti atoms induced by ferroelectric displacements. For the electrically-polarized multilayer ($m = 4$), the upward polarization makes Ti atoms to move away from the bottom interface and towards the top interface (Fig.1). This causes the Fe-Ti bond length to be shorter and hence the overlap between the Fe 3*d* and Ti 3*d* orbitals to be stronger at the top interface compared to the bottom interface. As a consequence, the minority-spin bonding state lies deeper in energy and hence is more populated for Fe and Ti atoms at the top interface than at the bottom interface, as is indicated by the red and blue arrows in Figs.2a,b. Thus, ferroelectric instability enhances the induced magnetic moment on top Ti atoms but reduces the magnetic moment of bottom Ti atoms. The change is opposite for Fe atoms: their magnetic moments are enhanced at the bottom interface but reduced at the top interface.

With increasing $BaTiO_3$ thickness, the net polarization of the ferroelectric film grows and gradually approaches its bulk value. This is evident from Fig. 3 revealing the increasing relative displacements between Ti and O atoms which saturate at the bulk value of 0.125 Å. (These displacements are inhomogeneous across the ferroelectric film reflecting an inhomogeneous local polarization similar to that we found for $Pt/KNbO_3$ multilayers.[16]) This enhances the asymmetry in the strength of the bonding and consequently in the magnetic moments at the top and bottom interfaces. As is seen from Table 1, with increasing thickness of the $BaTiO_3$ layer from $m = 4$ to $m = 16$ the difference in the Fe magnetic moments increases from $?\mu_{Fe} = 0.06\ \mu_B$ to $?\mu_{Fe} = 0.12\ \mu_B$, and the difference in the Ti magnetic moments increases from $?\mu_{Ti} = 0.22\ \mu_B$ to $?\mu_{Ti} = 0.25\ \mu_B$.

**Table 1.** Magnetic moments (in units of $\mu_B$) of Fe and Ti atoms at the top and bottom interfaces of $Fe/BaTiO_3$ multilayers. $\Delta \mu = \mu_{bot} - \mu_{top}$ is the difference between the magnetic moments at the top and bottom interfaces. The polarization of the $BaTiO_3$ film is pointing upward.

|  | $M = 2$ | $m = 4$ | $m = 6$ | $m = 8$ | $m = 10$ | $m = 16$ |
|---|---|---|---|---|---|---|
| $Fe_{top}$ | 2.59 | 2.61 | 2.60 | 2.59 | 2.58 | 2.56 |
| $Fe_{bot}$ | 2.59 | 2.67 | 2.67 | 2.67 | 2.67 | 2.68 |
| $\Delta \mu_{Fe}$ | 0.0 | 0.06 | 0.07 | 0.08 | 0.09 | 0.12 |
| $Ti_{top}$ | -0.30 | -0.40 | -0.40 | -0.40 | -0.40 | -0.40 |
| $Ti_{bot}$ | -0.30 | -0.18 | -0.17 | -0.16 | -0.16 | -0.15 |
| $\Delta \mu_{Ti}$ | 0.0 | 0.22 | 0.23 | 0.24 | 0.24 | 0.25 |

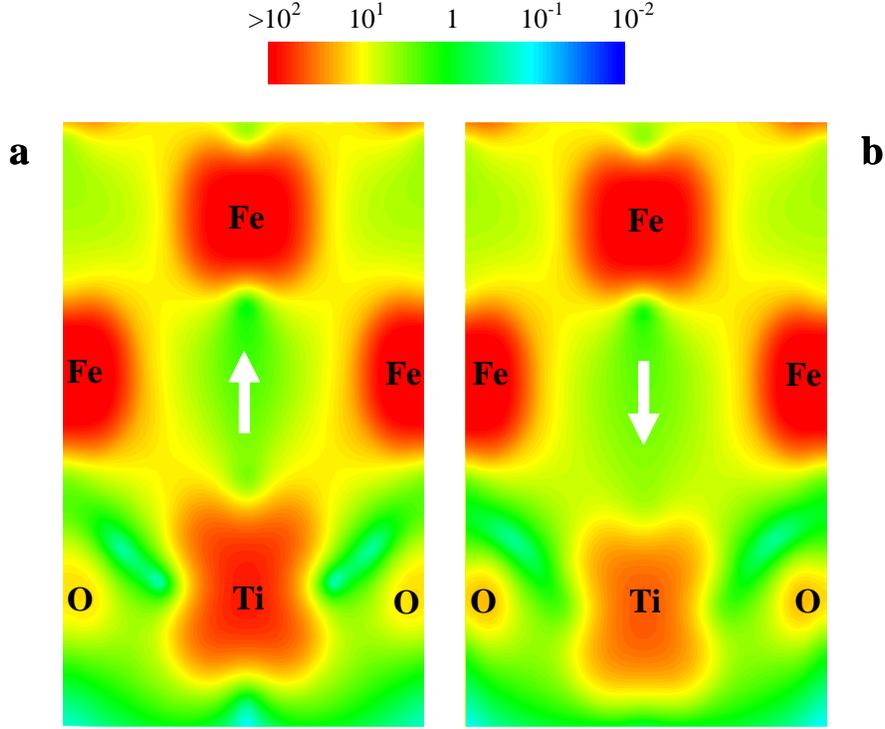

**Figure 4.** Minority-spin charge density (in arbitrary units) at the Fe/BaTiO$_3$ interface for two opposite polarizations in BaTiO$_3$. The charge density is calculated in the energy window from $E_F$ − 1 eV to $E_F$ in the (010) plane for the m = 4 multilayer. (a) Net polarization pointing up; (b) net polarization pointing down.

Different magnetic properties of the two interfaces reflect the change which occurs at *one* interface if the polarization in BaTiO$_3$ reverses. To illustrate this change and to obtain a further insight into the origin of the excess magnetic moments induced by ferroelectricity, we calculated the interface electronic charge density for two opposite polarization orientations in BaTiO$_3$. Fig.4 shows the minority-spin charge density calculated in the energy window from $E_F - 1\text{eV}$ to $E_F$ covering the region where the Fe-Ti minority-spin bonding states are located (see Figs.2a,b). As is seen from Figs.4a,b, the charge distribution at the interface Ti atom has a shape of the $d_{xz}$ ($d_{yz}$) orbital which is hybridized with the respective *d* orbitals on the nearest Fe atoms. It is evident from the comparison of the charge density shown in Figs.4a and 4b that the overlap between the Ti and Fe electronic clouds is much stronger for the local polarization pointing into the Fe film than for the local polarization pointing away from the Fe film, which reflects a stronger hybridization for the former. Since the majority-spin density on the interface Ti atoms is small for the energies from $E_F - 1\,\text{eV}$ to $E_F$ (see Fig.2a), the minority-spin density shown in Figs. 4a,b on the Ti atoms reveals the difference in the spin density on these atoms which is the source of the magnetic moment change for opposite polarizations in BaTiO$_3$.

The predicted results suggest a possibility to observe a *net* magnetization change in a Fe/BaTiO$_3$ *bilayer* deposited on a proper substrate. In this case, there is only *one* Fe/BaTiO$_3$ interface which has magnetic properties dependent on the orientation of the ferroelectric polarization, and hence the polarization reversal will inevitably change the magnetic moment of the entire system. We estimate the magnetoelectric coefficient ***a*** of this multiferroic bilayer by taking the ratio of the magnetization change $\Delta M$ to the coercive electric field $E_c$ of the BaTiO$_3$ film. Assuming that the Fe and BaTiO$_3$ layers have thicknesses of 1 nm and 2.5 nm ($m = 6$) respectively and taking the change in the interface magnetic moment per unit cell of 0.3 ***m*B** (as follows from our calculations for $m = 6$), we find that the average magnetization change in the Fe/BaTiO$_3$ bilayer is about ***m*$_0\Delta M \approx 120\,\text{G}$. Since coercive fields of BaTiO$_3$ films are in the range of $E_c \approx 10$ kV/cm,[17] we obtain ***a*** $\simeq$ ***m*$_0\Delta M / E_c \approx 0.01$ G cm/V which is of the same order in magnitude as the magnetoelectric coefficient measured in epitaxial BiFeO$_3$/CoFe$_2$O$_4$ columnar nanostructures.[10] Thus, the magnetoelectric effect induced by interface bonding in ferroelectric/ferromagnetic multilayers can be as large as the magnetoelectric effect induced by strain in vertically aligned structures. We note, however, that the predicted phenomenon is qualitatively different from the "standard" magnetoelectric effect which is the volume effect and for which the magnetization is a linear function of the applied electric field. In our case the magnetoelectric effect is confined to the interface and represents a change of the interface magnetic moment at the coercive field of the ferroelectric. Since this phenomenon is primarily due to the electronic hybridization between the transition metal elements with less than half occupied *d* bands (Ti) and more than half occupied *d* bands (Fe), any ferromagnetic/ferroelectric multilayer with such a combination of elements is predicted to have the magnetoelectric coefficient similar to that found for the Fe/BaTiO$_3$ system. We therefore hope that our theoretical predictions will stimulate experimental studies of such multilayers to search for the magnetoelectric effect driven by interface bonding.

This work was supported by NSF (grants Nos. DMR-0203359 and MRSEC DMR-0213808) and the Nebraska Research Initiative. Computations were performed utilizing the Research Computing Facility of the University of Nebraska-Lincoln.